\documentclass[11pt, a4paper]{article}

\usepackage{amsmath,amssymb}

\usepackage{graphicx}

\usepackage{color}

\numberwithin{equation}{section}
\newcommand{\p}{\partial}
\newcommand{\nn}{\nonumber}
\newcommand{\F}{\mathcal{F}}

\newcommand{\la}{\langle}
\newcommand{\ra}{\rangle}
\newcommand{\al}{\alpha}
\newcommand{\lm}{\lambda}

\DeclareMathOperator{\res}{Res}

\newtheorem{thm}{Theorem}[section]

\newtheorem{conj}[thm]{Conjecture}

\newtheorem{lem}[thm]{Lemma}

\newtheorem{rmk}[thm]{Remark}

\newenvironment{prf}{\noindent {\it Proof} \ }{\hfill $\Box$}

\begin{document}

\title{Proof of a Conjecture on the Genus Two Free Energy Associated to the $A_n$ Singularity}
\author{Yulong Fu, Si-Qi Liu, Youjin Zhang, Chunhui Zhou\\
{\small Department of Mathematical Sciences,
Tsinghua University}\\
{\small Beijing 100084, P. R. China}}
\date{}\maketitle

\begin{abstract}
In a recent paper \cite{DLZ}, it is proved  that the genus two free energy of an arbitrary semisimple Frobenius manifold can be represented as a sum of contributions associated with dual graphs of certain stable algebraic curves of genus two plus the so called genus two G-function, and  for a certain class of Frobenius manifolds it is conjectured that the associated genus two G-function vanishes. In this paper, we prove this conjecture for the Frobenius manifolds associated with simple singularities of type A.
\end{abstract}

\tableofcontents

\section{Introduction} \label{sec-1}

The notion of Frobenius is a geometrical characterization of the Witten-Dijkgraaf-Verlinde-Verlinde (WDVV) equations of associativity that arise in the study of 2D topological field theory (TFT)\cite{Dij1, Dij2, D1, Wit1}. For a 2D TFT with $n$ primary fields, the generating function of its correlators,  called the free energy, has the genus expansion
\begin{equation}\label{gexp}
\F({\bf t}) = \sum_{g\geq 0} \epsilon^{2g-2} \F_g({\bf t}),
\end{equation}
where for any $g\in \mathbb{Z}_{\ge 0}$ the function $\F_g({\bf t})$ is called the genus $g$ free energy, it is a function defined on the large phase space of the 2D TFT with coordinates
\[{\bf t} =\left( t^{\alpha, p}\right), \quad \alpha=1, \dots, n, \quad p=0, \, 1, \, 2, \dots.\]
The restriction of the genus zero free energy $\F_0({\bf t})$ yields a function
\[ F(v^1,\dots,v^n)=\left.\F_0({\bf t})\right|_{t^{\alpha,p}=0 (p>0),\ t^{\alpha,0}=v^\alpha}\]
of $n$ variables $v^1,\dots, v^n$ which satisfies the WDVV equations of associativity.
In terms of the corresponding Frobenius manifold, the function $F(v)$ is called the potential, and
the variables $v^1,\dots, v^n$ are called the flat coordinates of the Frobenius manifold w.r.t.
the flat metric defined by
\begin{equation}
\eta_{\alpha\beta}=\frac{\p^3 F(v)}{\p v^1\p v^\alpha\p v^\beta},\quad
\alpha,\beta=1,\dots,n.
\end{equation}

One of the important subjects of study in the theory of Frobenius manifold is to reconstruct
the full genera free energy $\F({\bf t})$ in terms of the geometric structure of the Frobenius manifold. The reconstruction of the genus zero free energy $\F_0({\bf t})$ is achieved with
the help of a particular solution
\begin{equation}
v({\bf t})=\left(v^1({\bf t}), \dots, v^n({\bf t})\right)
\end{equation}
of an integrable hierarchy of the form
\[\frac{\p v^\al}{\p t^{\beta,q}}= K^\alpha_{\beta,q;\gamma}(v) v^\gamma_x,\quad
\alpha, \beta=1,\dots, n,\ q\ge 0.\]
Here and in what follows summation w.r.t. repeated upper and lower greek indices is assumed.
This integrable hierachy is defined on the loop space of the Frobenius manifold, and is
called the Principal Hierarchy in \cite{DZ}. With the identification of the spatial variable $x$ with the time variable $t^{1,0}$, this particular solution of the integrable
hierarchy has the following relation with the genus zero free energy $\F_0({\bf t})$:
\[ v^\alpha({\bf t})= \eta^{\alpha\gamma}\frac{\p^2\F_0({\bf t})}{\p t^{1,0}\p t^{\gamma, 0}},\quad \alpha=1,\dots,n.\]
Here $(\eta^{\alpha\beta})=(\eta_{\alpha\beta})^{-1}$.

For the reconstruction of the higher genera free energies for any semisimple Frobenius
manifold, an algorithm is given in \cite{DZ} by solving recursively the so called loop equation of the Frobenius manifold. The genus $g (g\ge 1)$ free energy $\F_g({\bf t})$ can be represented
in the form
\begin{equation}\label{gexp1}
\F_g({\bf t})=\hat \F_g\left( v({\bf t}), v_x({\bf t}), \dots, v^{(3g-2)}({\bf t})\right), \end{equation}
where $v({\bf t})$ is the particular solution of the Principal Hierarchy, and 
\[v^{(k)}({\bf t})=(\p_x^k v^1({\bf t}), \dots, \p_x^k v^n({\bf t})).\]
In particular, the genus one free energy has the expression \cite{Dij3, DZ-cmp, Getz}
\begin{equation}\label{f1}
\F_1({\bf t})=\frac1{24}\log \det(c_{\alpha\beta\gamma}(v({\bf t}) v^\gamma_x({\bf t})))+G(v({\bf t})),
\end{equation}
where
\[c_{\al\beta\gamma}(v)=\frac{\p^3 F(v)}{\p v^\al\p v^\beta\p v^\gamma}\]
and $G(v)$ is called the genus one G-function of the semisimple Frobenius manifold.
The function $G(v)$ is defined by the isomonodromic tau function $\tau_I$ and the Jacobian $J$ of the
transformation between the flat coordinates $v^1,\dots, v^n$ and the canonical coordinates $u_1,\dots, u_n$ (see their meaning below) of the Frobenius manifold via the formula
\[G(v)=\log\frac{\tau_I(v)}{J^{1/{24}}(v)}.\]
Note that in the expression \eqref{f1} of the genus one free energy $\F_1$, the first
term in the r.h.s. of the formula can be represented by the genus zero three point
correlation functions since we have
\begin{equation}
c_{\alpha\beta\gamma}(v({\bf t})) v^\gamma_x({\bf t})=\frac{\p^3\F_0({\bf t})}{\p t^{1,0}\p t^{\al,0}\p t^{\beta,0}}\,.
\end{equation}
Thus the genus one free function can be represented as the summation of two parts,
the first part can be represented explicitly by using the genus zero correlation functions, and the second part is given by the G-function defined on the Frobenius manifold. In general one do not know the explicit expression of the G-function given in terms of the flat coordinates. However, it is proved in \cite{Hert, stra} that for a semisimple Frobenius manifold that is associated to a simple singularity or, equivalently, to the a Coxeter group of ADE type, the function $G(v)$ vanishes.

Similar to the above expression of the genus one free energy, it is shown in a recent paper \cite{DLZ} that the genus two free energy can also be represented as a summation of two parts: the first part is given in an explicit way by some genus zero correlation functions, and the second part is given by the so called genus two G-function.

\begin{thm}[\cite{DLZ}]  \label{main-thm}
Let $M$ be a semisimple Frobenius manifold of dimension $n$. Then the genus two free energy has the expression
\begin{equation}\label{mthm-zh}
\F_2=\sum_{p=1}^{16} c_p\,Q_p+G^{(2)}(u, u_x, u_{xx}).
\end{equation}
Here each term $Q_p$ corresponds to a dual graph of a stable curve of arithmetic genus two and can be represented by some genus zero correlation functions; $c_1,\dots, c_{16}$ are some constants.
The function $G^{(2)}(u, u_x, u_{xx})$ is called the \emph{genus two $G$-function} of the Frobenius manifold, and has an explicit expression \eqref{g2-zh}
represented in terms of the canonical coordinates $u_1$, \dots, $u_n$ of the Frobenius manifold.
\end{thm}

In \cite{DLZ} it is also conjecture that for a certain class of Frobenius manifolds the genus two G-functions equal to zero.
\begin{conj}\label{main-conj1}
If $M$ is a Frobenius manifold associated to an ADE singularity or an extended affine Weyl groups of ADE type, then
\begin{equation}
G^{(2)}(u, u_x, u_{xx})=0.
\end{equation}
\end{conj}
The construction of the two classes of Frobenius manifold structures mentioned in the above conjecture can be found in \cite{D1, dz-comp, saito}. They can also be interpreted in terms of 
cohomological field theory and Gromov-Witten invariants of $\mathbb{P}^1$-orbifolds,
see \cite{fsz, for-1, for, milanov, rossi, taka} and references therein.

The purpose of the present paper is to prove the following theorem.
\begin{thm}\label{mth}
For the class of Frobenius manifolds obtained from the simple singularities of type A, the above conjecture holds true.
\end{thm}

The paper is organized as follows. In Section \ref{sec-2} we represent the rotation and Lam\'e coefficients of the Frobenius manifolds associated to the simple singularities of type A in terms of their superpotentials. In Section \ref{sec-3} we
prove some identities that will be used in the subsequent sections. 
In Section \ref{sec-4} we prove the vanishing of the coefficients $G_i^{(2)}(u)$ and $G_{ij}^{(2)}(u)$ that appear in the expression of the genus two G-function. In Section \ref{sec-4}--\ref{sec-6} we finish the proof of Theorem \ref{mth}.  
Section \ref{sec-7} is an conclusion of the paper.

\section{The Rotation and Lam\'e Coefficients} \label{sec-2}
Let us recall the definition of the rotation coefficients and Lam\'e coefficients of a semisimple Frobenius manifold $(M^n, \cdot\,, \la\,,\,\ra, e, E)$ . Near each point of the Frobenius manifold there is a system of local coordinates $u_1,\dots, u_n$ given by the roots of the characteristic polynomial of the operator of multiplication by the Euler vector $E$. They are called the canonical coordinates of the Frobenius manifold. In these coordinates the multiplication table defined on the tangent space
of $M$ is given by
\[\frac{\p}{\p u_i}\cdot \frac{\p}{\p u_j}=\delta_{ij}\,\frac{\p}{\p u_i},\quad i, j=1,\dots, n.\]
In the canonical coordinates the unity vector field $e$ and the Euler vector field $E$ have the expressions
\[e=\sum_{i=1}^n \frac{\p}{\p u_i},\quad E=\sum_{i=1}^n u_i\,\frac{\p}{\p u_i}\,,\]
and the flat metric $\la\,,\,\ra$ of the Frobenius manifold takes the diagonal form
\[ \sum_{i=1}^n \eta_{ii}(u) du_i^2\, .\]
The Lam\'e coefficients $h_i$ and the rotation coefficients $\gamma_{ij}$ of the above diagonal metric are defined by
\[ h_i=h_i(u)=\sqrt{\eta_{ii}},\quad \quad i=1,\dots,n\]
and
\[\gamma_{ij}=\gamma_{ji}=\frac1{h_i}\frac{\p h_j}{\p u_i}\ \ \textrm{for}\ i\ne j,\quad \gamma_{ii}=0\]
for some choice of the signs of the square roots. They satisfy the following equations:
\begin{align}
&\frac{\p h_i}{\p u_k}=\gamma_{ik} h_k \ \ \textrm{for}\ i\ne k,\quad
\frac{\p h_i}{\p u_i}=-\sum_{k=1}^n \gamma_{ik} h_k,\label{zh-5}\\
&\frac{\p\gamma_{ij}}{\p u_k}=\gamma_{ik}\gamma_{kj}\ \ \textrm{for}\
k\ne i, j,\quad  \frac{\p\gamma_{ij}}{\p u_i}=\frac{\sum_{k=1}^n(u_j-u_k)\gamma_{ik}\gamma_{kj}-\gamma_{ij}}{u_i-u_j}.\label{zh-6}
\end{align}

Now let us consider semisimple Frobenius manifold associated to the singularity
$f(z)=z^{n+1}$ of type $A_n$. The miniversal unfolding of the function $f(z)$
is given by
\begin{equation}
\lm(z,t)=z^{n+1}+t_n z^{n-1}+\dots+t_1,\quad (z,t)\in\mathbb{C}\times B,
\end{equation}
where $B$ is an open ball in $\mathbb{C}^n$. Denote by $C\subset B$ the caustic
and $M=B\setminus C$. Then on $M$ there is a semisimple Frobenius manifold structure
given by the flat metric
\begin{equation}\label{residue-pairing}
\la \p', \p''\ra_t = -\res_{z=\infty}\frac{(\p' \lm(z,t))(\p'' \lm(z,t))\,dz}{\p_{z} \lm(z,t)} \end{equation}
and the multiplication
\begin{equation}\label{residue-pairing}
\la \p'\cdot \p'', \p'''\ra_t = -\res_{z=\infty}\frac{(\p' \lm(z,t))(\p'' \lm(z,t))(\p''' \lm(z,t))\,dz}{\p_{z} \lm(z,t)}
\end{equation}
for any $\p',\ \p'', \p'''\in T_t M$. The unity vector field and the Euler vector field are
defined by
\begin{equation}
e=\frac{\p}{\p t_1},\quad E=\sum_{\al=1}^n \frac{n+2-\al}{n+1} t_\al \frac{\p}{\p t_\al}.
\end{equation}
Note that the flat coordinates $v^1,\dots, v^n$ of the
metric $\la\,,\,\ra$ can be represented as quasihomogenius polynomials of  $t_1,\dots, t_n$ by the formula
\[v^\al=-\frac{n+1}{n+1-\al} \res_{z=\infty}\lm(z)^{\frac{n+1-\al}{n+1}} dz,\quad \al=1,\dots,n.\]
For example, when $n=1,2,3$ the corresponding Frobenius manifolds have the following potentials respectively:
\begin{align*}
&F(v)=\frac1{12} (v^1)^3,\quad v^1=t_1.\\
&F(v)=\frac16 (v^1)^2 v^2-\frac1{216} (v^2)^4,\quad v^1=t_1,\ v^2=t_2.\\
&F(v)=\frac18 (v^1)^2 v^3+\frac18 v^1 (v^2)^2-\frac1{64}(v^2)^2 (v^3)^2+ \frac1{3840} (v^3)^5,\\
&\qquad \quad v^1=t_1-\frac18 t_3^2,\ v^2=t_2,\ v_3=t_3.
\end{align*}

Let $z_1,\dots, z_n$ be the critical points of $\lm(z,t)$ satisfying
\[ \p_z\lm(z,t) |_{z=z_i}=0,\quad i=1,\dots,n.\]
Then  the canonical coordinates of the Frobenius manifold are given by the
the critical values
\begin{equation}
u_i(t)=\lm(z_i,t),\quad i=1,\dots,n.
\end{equation}
The Lam\'e coefficients $h_i$ and the rotation coefficients $\gamma_{ij}$ can be represented as follow \cite{DLZ}:
\begin{equation}\label{ga-A}
h_i=\frac1{\sqrt{\lm''(z_i,t)}},\quad
\gamma_{ij}=\frac{h_i\,h_j}{(z_{i}-z_{j})^2},\quad i, j=1,\dots, n.
\end{equation}
Here and in what follows we denote $\lm(z,t)$ by $\lm(z)$, and derivatives of the function $\lm(z)$ are taken w.r.t. the variable $z$.

To simplify the expressions that will
be given in what follows, we introduce the notations
\begin{equation}\label{zh-5-10}
H_i=\frac12\sum_{j\ne i} u_{ij}\gamma_{i j}^2,\quad 1\le i\le n
\end{equation}
with $u_{ij}=u_i-u_j$, and
\begin{equation}\label{zh-17}
z_{ij}=z_i-z_j,\quad C_{ik}=\frac{\lm^{(k)}(z_i)}{\lm''(z_i)},\quad
i, k=1,\dots,n.
\end{equation}
The following lemma lists some identities that will be used frequently in the proof of Theorem \ref{mth}.
\begin{lem}\label{zh-16}
The Lam\'e coefficients $h_i$ and the rotation coefficients $\gamma_{ij}$ satisfy the following identities:
\begin{align}
\partial_kh_i&=\frac{h_ih_k^2}{z_{ik}^2}\quad \textrm{for}\ k\neq i;\quad
\partial_ih_i=h_i^3\left(\frac14{C_{i3}^2}-\frac16 C_{i4}\right);\label{zh-1}\\
\partial_k\gamma_{ij}&=\frac{h_ih_jh_k^2}{z_{ik}^2z_{jk}^2}\quad \textrm{for}\
\textrm{distinct}\ i, j, k;\label{zh-2}\\
\partial_i\gamma_{ij}&=\frac{h_i^3h_j}{z_{ij}^2}\left(\frac{3}{z_{ij}^2}+\frac{C_{i3}}{z_{ij}}+\frac{C_{i3}^2}{4}-\frac{C_{i4}}{6}\right);\label{zh-3}\\
H_i&=\frac1{48}{h_i^2}\left(C_{i3}^2-C_{i4}\right).\label{zh-11}
\end{align}
Here $\p_k=\frac{\p}{\p u_k}$.
\end{lem}
\begin{prf}
The lemma can be proved by using the identities \eqref{zh-5}, \eqref{zh-6}, the
formulae \eqref{ga-A} and the residue theorem on the complex plane.
Let us show the details of the derivation for the identity \eqref{zh-2} as follows:
\begin{align*}
\partial_i\gamma_{ij}&=\frac{1}{u_{ij}}\left(\sum_{k\neq i,j}u_{jk}\gamma_{ik}\gamma_{kj}-\gamma_{ij}\right)=\frac{h_i h_j}{u_{ij}}\sum_{k\neq i,j}\frac{u_j-u_k}{z_{ik}^2 z_{jk}^2\lambda{''}(z_k)}-\frac{\gamma_{ij}}{u_{ij}}\\
&=\frac{h_i h_j}{u_{ij}}\sum_{k\neq i,j}\res_{z=z_k}\frac{\lambda(z_j)-\lambda(z)}{(z-z_i)^2 (z-z_j)^2 \lambda^{'}(z)}-\frac{\gamma_{ij}}{u_{ij}}\\
&=\frac{h_i h_j}{u_{ij}}\Big(\res_{z=z_i}+\res_{z=z_j}\Big)\frac{\lambda(z)-\lambda(z_j)}{(z-z_i)^2 (z-z_j)^2 \lambda^{'}(z)}-\frac{\gamma_{ij}}{u_{ij}}\\
&=\frac{h_i^3h_j}{z_{ij}^2}\left(\frac{3}{z_{ij}^2}+\frac{C_{i3}}{z_{ij}}+\frac{C_{i3}^2}{4}-\frac{C_{i4}}{6}\right).
\end{align*}
Here the last equality is due to a residue formula for $R_6(2,2)$ given in appendix B.  
The lemma is proved.
\end{prf}
\begin{rmk}
In what follows we will frequently calculate residues of some rational functions, so for 
the readers convenience we list some useful residue formulae in Appendix B.
\end{rmk}

\section{Some useful identities}\label{sec-3}
In this section, we list some identities that will be used in the subsequent sections to prove 
the main theorem of the present paper.

For fixed distinct indices $i, k \in\{1,2,\dots, n\}$, let us denote
 \[{A}_{i,k,p}:=\sum_{j\neq k,i}\frac{1}{z_{kj}^p}\,.\]
In order to simplify the notations we will write $A_{i,k, p}$ by $A_p$ in what follows, keeping in mind that the indices $i, k$ are fixed. By using identities among symmetric polynomials we have
\begin{align}
&\sum_{\substack{j_1<\ldots<j_p\\j_1,\ldots,j_p\neq i, k}}\frac{1}{z_{kj_1}\ldots z_{kj_p}}=\notag \\
&\begin{cases}
\dfrac{1}{2}(A_1^2-A_2), & \text{p=2;}\\[10pt]
\dfrac{1}{6}(A_1^3-3A_1A_2+2A_3), & \text{p=3;}\\[10pt]
\dfrac{1}{24}(A_1^4-6A_1^2A_2+8A_1A_3+3A_2^2-6A_4), & \text{p=4;}\\[10pt]
\dfrac{1}{120}(A_1^5-10A_1^3A_2+20A_1^2A_3+15A_1A_2^2\\ \qquad  \ -30A_1A_4-20A_2A_3+24A_5), & \text{p=5;}\\[10 pt]
\dfrac{1}{720}(A_1^6-15A_1^4A_2+40 A_1^3 A_3+45A_1^2A_2^2\\
\qquad-90A_1^2A_4+144A_1A_5-120 A_1A_2 A_3\\
\qquad\quad  -15A_2^3+90A_2A_4+40A_3^2
-120A_6), & \text{p=6}.\label{l2}
\end{cases}
\end{align}
These identities enable us to represent the rational functions $z_{ik}^pC_{k,p+2}$
in terms of $A_p$ by using the following relation:
\begin{align}
&z_{ik}^p C_{k,p+2}=(p+1)!z_{ik}^p\sum_{\substack{j_1<\ldots<j_p\\j_1,\ldots,j_p\neq k}}\frac{1}{z_{kj_1}\ldots z_{kj_p}}\label{zh-5-7}\\
&=(p+1)!\left(z_{ik}^p\sum_{\substack{j_1<\ldots<j_p\\j_1,\ldots,j_p\neq i, k}}\frac{1}{z_{kj_1}\ldots z_{kj_p}}-z_{ik}^{p-1}\sum_{\substack{j_1<\ldots<j_{p-1}\\j_1,\ldots,j_{p-1}\neq i, k}}\frac{1}{z_{kj_1}\ldots z_{kj_{p-1}}}\right).\notag
\end{align}
For example, when $p=1,2$ we have the following identities respectively:
\begin{align}
&z_{ik}C_{k3}=2\big(-1+{A}_1z_{ik}\big),\label{zh-13} \\
&z_{ik}^2C_{k4}= 3\left({A}_1^2-{A}_2\right) z_{ik}^2-6 A_1 z_{ik}.\label{zh-14}
\end{align}
We can also represent the rational function $\frac{h_i^2}{h_i^2}$ in terms of $A_p$ 
due to the relation
\begin{equation}
\frac{h_k^2}{h_i^2}=-\prod_{j\neq k,i}\frac{z_{ij}}{z_k-z_j}
 =-1-\sum_{p=1}^{n-2}\left(\sum_{\substack{j_1<\ldots<j_p\\j_1,\ldots,j_p\neq k,i}}\frac{1}{z_{kj_1}\ldots z_{kj_p}}\right)z_{ik}^p. \label{zh-12} \\
\end{equation}

By using the above identities, we arrive at the following lemma.
\begin{lem}\label{lem-2}
For any fixed $i\in\{1,2,\dots,n\}$ we have the following four identities:
\begin{align}
&\sum_{l\neq i}\frac{h_l^2}{z_{il}^2}\left(C_{l3}^2-C_{l4}-\frac{2C_{l3}}{z_{il}}\right)\label{iden-1} \\
&=-\frac{h_i^2}{12}\left(6C_{i3}^4-15C_{i3}^2C_{i4}+4C_{i4}^2+7C_{i3}C_{i5}-2C_{i6}\right);\notag
\\ \notag
&\\
&\sum_{l\neq i}\frac{h_l^4}{z_{il}^2}\left(\frac{3}{z_{il}^2}-\frac{C_{l3}}{z_{il}}\right)\label{iden-2}\\
&=\frac{h_i^4}{240}(75C_{i3}^4-120C_{i3}^2C_{i4}+20C_{i4}^2+30C_{i3}C_{i5}-4C_{i6});\notag\\
&\notag \\
&\sum_{l\neq i}\frac{h_l^2}{z_{il}^3}\left(C_{l3}^2-C_{l4}-\frac{3C_{l3}}{z_{il}}\right)\label{iden-3}\\
&=-\frac{h_i^2}{240}(75C_{i3}^5-240C_{i3}^3C_{i4}+140C_{i3}C_{i4}^2+120C_{i3}^2C_{i5}-60C_{i4}C_{i5}\notag\\
&\quad -44C_{i3}C_{i6}+10C_{i7});\notag\\
&\notag \\
&\sum_{l\neq i}\frac{h_l^2}{z_{il}^4}\left(C_{l3}^2-C_{l4}-\frac{4C_{l3}}{z_{il}}\right)\label{iden-4}\\
&=-\frac{h_i^2}{720}(135C_{i3}^6-525C_{i3}^4C_{i4}+480C_{i3}^2C_{i4}^2-60C_{i4}^3+270C_{i3}^3C_{i5}\notag\\
&\quad -300C_{i3}C_{i4}C_{i5}+30C_{i5}^2-108C_{i3}^2C_{i6}+52C_{i4}C_{i6}+32C_{i3}C_{i7}-6C_{i8}).\notag
\end{align}
\end{lem} 
\begin{prf} Let us take the first identity as an example to illustrate the proof of the lemma. Regarding $z_i$ as an independent variable, we denote the functions defined
by the r.h.s and the l.h.s. of \eqref{iden-1} by $f_1(z_i)$ and $f_2(z_i)$ respectively.
Since both of these rational functions tend to zero when $z_i$ tends to infinity, in order
to prove the identy \eqref{iden-1} we 
only need to show that these two functions have identical principal parts at each of their 
poles $z_k (k\neq i)$. Note that the orders of these poles do not exceed $5$
and $z_i=z_k$ is not a zero or a pole of the function,
so it suffices to show that the Taylor expansions of the functions $z_{ik}^4 f_1(z_i)/h_i^2$ and $z_{ik}^4f_2(z_i)/h_i^2$ in $z_i-z_k$ coincide up to the terms of order $4$. 

It follows from the identities \eqref{l2}--\eqref{zh-12} that 
\begin{align}
&z_{ik}^4 \frac{f_1(z_i)}{h_i^2}=-8-2 A_1 z_{ik}+(A_1^2+A_2) z_{ik}^2+
\frac23 \left(A_1^3-3 A_1 A_2-4 A_3\right) z_{ik}^3\notag\\
&\qquad +\frac16\left(A_1^4-12 A_1^2 A_2+3 A_2^2-4 A_1 A_3+12 A_4\right) z_{ik}^4+{\cal O}(z_{ik}^5).\label{zh-5-8}
\end{align}

By using the identities
\[\sum_{j\neq k,i}\frac{1}{z_{ij}^p}=\sum_{m\ge 0} (-1)^m \binom{-p}{m} A_{p+m} \,z_{ik}^m,\quad p\ge 1\]
and \eqref{zh-5-7} we can also represent the functions $z_{ki}^p C_{i, p+2}$ in terms of 
the functions $A_1, A_2,\dots$. Thus we can obtain the Taylor expansion of the function
$z_{ik}^4f_2(z_i)/h_i^2$ at $z_k$, and it turns out that its Taylor expansion coincide 
with \eqref{zh-5-8} up to the $z_{ik}^4=(z_i-z_k)^4$ term. The lemma is proved.
\end{prf}

\section{Vanshing of $G_{i}^{(2)}$ and $G_{ij}^{(2)}$}\label{sec-4}
Let us fix $k\neq i$ and regard $\dfrac{u_{k,x}}{u_{i,x}}(k\neq i)$ as an independent variable in the expression of $G_i^{(2)}$ given in Appendix A, 
then the sum of all its coefficients that appear in the expression of $G_i^{(2)}$ is given by 
\begin{equation}\begin{split}
T(\dfrac{u_{k,x}}{u_{i,x}}):=&\frac{\partial_kh_iH_i}{60h_i^3}-\frac{7\partial_ih_i\partial_kh_i}{5760h_i^4}+\frac{\gamma_{ik}H_k}{120h_ih_k}-\sum_l\frac{\gamma_{il}\partial_kh_i}{5760h_i^2h_l}\\
&-\frac{\gamma_{ik}\partial_kh_k}{1152h_ih_k^2}+\frac{\partial_i\gamma_{ik}h_k}{1920h_i^3}+\sum_{l\neq k,i}\frac{\partial_k\gamma_{il}}{5760h_ih_l}+\frac{\partial_k\gamma_{ik}}{5760h_ih_k}\\
&+\frac{\partial_k\gamma_{ik}}{2880h_ih_k}-\frac{7\gamma_{ik}^2}{1152h_i^2}
-\sum_{l\ne k, i}\frac{h_k\gamma_{il}\gamma_{kl}}{1920h_ih_l^2}\label{l4}
\end{split}\end{equation}

By using the formulae given in \eqref{ga-A} and in Lemma \ref{zh-16} we obtain the identities
\begin{align*}
\sum_l\frac{\gamma_{il}\partial_kh_i}{5760h_i^2h_l}&=\frac{h_k^2}{5760z_{ik}^2}\sum_{l\neq i}\frac{1}{z_{il}^2}
=\frac{h_k^2}{5760z_{ik}^2}\sum_{l\neq i}\res_{z=z_l}\frac{\lambda''(z)}{(z-z_i)^2\lambda'(z)}\\
&=-\frac{h_k^2}{5760z_{ik}^2}\res_{z=z_i}\frac{\lambda''(z)}{(z-z_i)^2\lambda'(z)}
=\frac{h_k^2}{5760z_{ik}^2}\big(\frac{C_{i3}^2}{4}-\frac{C_{i4}}{3}\big),
\end{align*}
and 
\begin{align*}
\sum_{l\neq k,i}\frac{\partial_k\gamma_{il}}{5760h_ih_l}
&=\frac{h_k^2}{5760z_{ik}^2}\sum_{l\neq i,k}\frac{1}{z_{kl}^2}
=\frac{h_k^2}{5760z_{ik}^2}(\sum_{l\neq k}\frac{1}{z_{kl}^2}-\frac{1}{z_{ik}^2})\\
&=\frac{h_k^2}{5760z_{ik}^2}\big(\frac{C_{k3}^2}{4}-\frac{C_{k4}}{3}-\frac{1}{z_{ik}^2}\big).
\end{align*}
Here we used the residue formula for $R_{1}(2)$ given in Appendix B. In a similar way, by using \eqref{ga-A} and the residue formula for $R_5(2,2)$ given in Appendix B we obtain the identity
\begin{align*}
\sum_l\frac{h_k\gamma_{il}\gamma_{kl}}{1920h_ih_l^2}&=\frac{h_k^2}{1920}\sum_{l\neq i,k}\frac{1}{z_{il}^2z_{kl}^2}
=\frac{h_k^2}{1920}\sum_{l\neq i,k}\res_{z=z_l}\frac{\lambda''(z)}{(z-z_i)^2(z-z_k)^2\lambda'(z)}\\
&=-\frac{h_k^2}{1920}(\res_{z=z_i}+\res_{z=z_k})\Big(\frac{\lambda''(z)}{(z-z_i)^2(z-z_k)^2\lambda'(z)}\Big)\\
&=-\frac{h_k^2}{1920z_{ik}^2}\Big(\frac{6}{z_{ik}^2}-\frac{C_{i3}-C_{k3}}{z_{ik}}-\frac{C_{i3}^2}{4}+\frac{C_{i4}}{3}-\frac{C_{k3}^2}{4}+\frac{C_{k4}}{3}\Big).
\end{align*}
Now the vanishing of $T(\dfrac{u_{k,x}}{u_{i,x}})$ easily follows from the above three identities. Thus by using the identities \eqref{zh-1}--\eqref{zh-11} we can represent the function $G_i^{(2)}$ in the form
\begin{align*}
G_i^{(2)}&=\frac{h_i^2}{5760}\left(\frac{3}{16}C_{i3}^4+\frac{1}{3}C_{i3}^2C_{i4}-\frac{11}{36}C_{i4}^2\right)\\
&+\sum_{k\neq i}\left(\frac{7h_k^2}{5760z_{ik}^2}\left(C_{k3}^2-C_{k4}-\frac{2C_{k3}}{z_{ik}}\right)
-\frac{h_k^4}{384h_i^2z_{ik}^2}\left(\frac{3}{z_{ik}^2}-\frac{C_{k3}}{z_{ik}}\right)\right)\\
&+\sum_{k\neq i}\left(\frac{h_i^2-11h_k^2}{576z_{ik}^4}+\frac{h_i^2C_{i3}}{1920z_{ik}^3}
+\frac{h_i^2}{5760z_{ik}^2}\left(\frac{C_{i3}^2}{2}-\frac{2C_{i4}}{3}+\frac{C_{k4}}{3}\right)\right).
\end{align*}
To simplify the above expression of $G_i^{(2)}$, we need to use the following identities:
\begin{align}
\sum_{k\neq i}\frac{1}{z_{ik}^p}&=\sum_{k\neq i}\res_{z=z_k}\frac{\lambda''(z)}{(z_i-z)^p\lambda'(z)}=-\res_{z=z_i}\frac{\lambda''(z)}{(z_i-z)^p\lambda'(z)},\label{l5}\\
\sum_{k\neq i}\frac{h_k^2}{z_{ik}^p}&=\sum_{k\neq i}\res_{z=z_k}\frac{1}{(z_i-z)^p\lambda'(z)}=-\res_{z=z_i}\frac{1}{(z_i-z)^p\lambda'(z)}\label{l6},\\
\sum_{k\neq i}\frac{C_{kq}}{z_{ik}^p}&=\sum_{k\neq i}\res_{z=z_k}\frac{\lambda^{(q)}(z)}{(z_i-z)^p\lambda'(z)}=-\res_{z=z_i}\frac{\lambda^{(q)}(z)}{(z_i-z)^p\lambda'(z)}.\label{l7}
\end{align}
Together with the residue formulae for $R_1(p), R_2(p)$ and $R_4(p,q)$ that are given in Appendix B, the above identities enable us to simplify the expression of $G_i^{(2)}$ further to obtain
\[ G_i^{(2)}=f_1(z_i)- f_2(z_i),\]
where
\[
f_1(z_i)=\sum_{l\neq i}\left(\frac{7h_l^2}{5760 \,z_{il}^2}\left(C_{l3}^2-C_{l4}-\frac{2C_{l3}}{z_{il}}\right)
-\frac{h_l^4}{384h_i^2 z_{il}^2}\left(\frac{3}{z_{il}^2}-\frac{C_{l3}}{z_{il}}\right)\right),\]
\[
f_2(z_i)=-\frac{h_i^2}{5760}\left(\frac{131}{16}C_{i3}^4-\frac{65}{4}C_{i3}^2C_{i4}+\frac{43}{12}C_{i4}^2+\frac{143}{24}C_{i3}C_{i5}-\frac{17}{12}C_{i6}\right).
\]
From the identities \eqref{iden-1}, \eqref{iden-2} it follows that $f_1(z_i)=f_2(z_i)$. Thus we proved that $G_i^{(2)}$ equals zero.

Now let us proceed to prove the vanishing of $G_{ij}^{(2)}$. Since $\gamma_{ii}=0$, we only need to show that $G_{ij}^{(2)}=0$ for $i\ne j$.  By using the formulae given in  \eqref{ga-A} and Lemma \ref{zh-16} we obtain
\begin{equation}\label{zh-15}
G_{ij}^{(2)}=\frac{h_i^2h_j^2}{5760z_{ij}^4}\left(\frac{6}{z_{ij}^2}-\frac{C_{i3}-C_{j3}}{z_{ij}}-\frac{1}{2}C_{j3}^2
+\frac{2}{3}C_{j4}\right)+\frac{h_i^2h_j^2}{2880z_{ij}^3}\sum_{k\neq i,j}\frac{1}{z_{ik}z_{jk}^2}.
\end{equation}
Then the vanishing of $G_{ij}^{(2)}$ follows from the following residue formula:
\begin{displaymath}\begin{split}
\sum_{k\neq i,j}\frac{1}{z_{ik}z_{jk}^2}&=-\sum_{k\neq i,j}\res_{z=z_k}\frac{\lambda''(z)}{(z-z_i)(z-z_j)^2 \lm'(z)}\\
&=(\res_{z=z_i}+\res_{z=z_j})\frac{\lambda''(z)}{(z-z_i)(z-z_j)^2 \lm'(z)}\\
&=-\frac{3}{z_{ij}^3}+\frac{C_{i3}-C_{j3}}{2z_{ij}^2}+\frac{1}{z_{ij}}\left(\frac{C_{j3}^2}{4}-\frac{C_{j4}}{3}\right).
\end{split}\end{displaymath}
Thus we have proved that in the expression \eqref{g2-zh} of the function $G^{(2)}$,
the first two terms on the r.h.s. do not do not give any contribution. In the next two sections, we are to show that the sum of the remaining two terms also equals to zero.   

\section{The skew symmetry property of $P_{ij}^{(2)}$ for $i\neq j$}\label{sec-5}
In this section we are to  prove that the sum 
\[\frac12 \sum_{i\ne j} P_{ij}^{(2)}(u) u^i_x u^j_x \]
has no contribution to the function $G^{(2)}$, i.e. we need to show that
\[P_{ij}^{(2)}=-P_{ji}^{(2)},\quad i\ne j.\]  

By using \eqref{ga-A} and \eqref{zh-17} we easily obtain the following identity: 
\begin{align*}
&\sum_{k,l}\left(\frac{h_ih_j\gamma_{il}\gamma_{jl}}{h_kh_l^2}\left(\frac{\gamma_{kl}}{h_l}-\frac{\gamma_{jk}}{2h_j}\right)
-\frac{h_i\gamma_{ij}\gamma_{jl}\gamma_{kl}}{h_kh_l^2}\right)\\
&=\sum_{k\neq i,j}\frac{h_i^2h_j^2}{z_{ik}^2z_{jk}^2}\left(\frac{C_{k3}^2}{4}-\frac{C_{k4}}{3}
-\frac{C_{j3}^2}{8}+\frac{C_{j4}}{6}\right)
-\sum_{k\neq j}\frac{h_i^2h_j^2}{z_{ij}^2z_{jk}^2}\left(\frac{C_{k3}^2}{4}-\frac{C_{k4}}{3}\right).
\end{align*}
Then by using Lemma \eqref{zh-16} we can simplify $P_{ij}^{(2)}$ to arrive at
the following expression of it:
\begin{align*}
&\frac{h_i^2h_j^2}{z_{ij}^2}\left(\frac{41}{480z_{ij}^4}-\frac{41C_{i3}}{1440z_{ij}^3}
+\frac{41(-3C_{i3}^2+2C_{i4}+2C_{j4})}{17280z_{ij}^2}+\frac{C_{i3}(-3C_{j3}^2+4C_{j4})}{17280z_{ij}}\right.\\
&\quad \left.+\frac{C_{j3}^2(9C_{j3}^2-30C_{j4})-16C_{j4}(C_{i4}-C_{j4})+C_{i3}^2(-9C_{j3}^2+24C_{j4})}{207360}\right)\\
&+\sum_{k\neq i}\left(-\frac{h_i^2h_j^2}{z_{ij}^2z_{ik}^2}\Big(\frac{1}{160z_{ij}^2}+\frac{C_{i3}}{480z_{ij}}
+\frac{15C_{i3}^2-10C_{i4}+2C_{j4}}{17280}\Big)
-\frac{11h_j^2h_k^2}{480z_{ij}^2z_{ik}^4}\right.\\
&\quad \left.
-\frac{h_j^2h_k^2C_{i3}}{360z_{ij}^2z_{ik}^3}
-\frac{h_j^2h_k^2(3C_{i3}^2-2C_{i4})}{4320z_{ij}^2z_{ik}^2}
-\frac{h_j^2h_j^2}{240z_{ij}^2z_{ik}^4}-\frac{h_j^2h_j^2C_{i3}}{720z_{ij}^2z_{ik}^3}\right)\\
&+\sum_{k\neq j}\left(\frac{h_i^2h_j^2}{z_{ij}^2z_{jk}^2}\Big(\frac{1}{96z_{ij}^2}+\frac{C_{i4}}{1440z_{ij}}
+\frac{3C_{i3}^2+12C_{j3}^2-2C_{i4}-6C_{j4}}{17280}\Big)\right.\\
&\quad \left.+\frac{h_i^2h_j^2C_{k4}}{4320z_{ij}^2z_{jk}^2}+\frac{7h_i^2h_k^2}{160z_{ij}^2z_{jk}^4}\right)\\
&+\sum_{k\neq i,j}\left(\frac{11h_i^2h_j^2}{2880z_{ij}^2z_{ik}^2z_{jk}^2}
+\frac{h_i^2h_j^2C_{k3}}{480z_{ik}^2z_{jk}^3}-\frac{h_i^2h_j^2}{160z_{ik}^2z_{jk}^4}+\frac{h_i^2h_j^2}{96z_{ik}^4z_{jk}^2}\right)\\
&+\frac{7}{2880z_{ij}^2}\left(\sum_{k\neq i}\frac{2h_j^2h_k^2C_{k3}}{z_{ik}^3}-\sum_{k\neq j}\frac{h_i^2h_k^2(C_{k3}^2-C_{k4})}{z_{jk}^2}\right).
\end{align*}
We need the following residue formula to calculate the summations in the above expression of $P^{(2)}_{ij}$:
\begin{align*}
&\sum_{k\neq i,j}\frac{C_{k3}}{z_{ik}^2z_{jk}^3}=\sum_{k\neq i,j}\frac{\lambda'''(z)}{(z_i-z)^2(z_j-z)^3\lambda'(z)}\notag \\
&=-\big(\res_{z=z_i}+\res_{z=z_j}\big)\left(\frac{\lambda'''(z)}{(z_i-z)^2(z_j-z)^3\lambda'(z)}\right)\\
&=\frac{6C_{i3}+4C_{j3}}{z_{ij}^5}+\frac{3C_{i3}^2-3C_{j3}^2-6C_{i4}+6C_{j4}}{2z_{ij}^4}\\
&\quad +\frac{3C_{i3}^3+6C_{j3}^3-8C_{i3}C_{i4}-16C_{j3}C_{j4}+6C_{i5}+12C_{j5}}{12z_{ij}^3}\\
&\quad -\frac{3C_{j3}^4-10C_{j3}^2C_{j4}+4(C_{j4})^2+7C_{j3}C_{j5}-4C_{j6}}{24z_{ij}^2}.
\end{align*}
Similar identities can be obtained by applying the residue formulae for $R_5(p,q)$
given in Appendix B  to the r.h.s. of the following formula:
\begin{align}
\sum_{k\neq i,j}\frac{1}{z_{ik}^pz_{jk}^q}=&\sum_{k\neq i,j}\res_{z=z_k}\frac{\lambda''(z)}{(z_i-z)^p(z_j-z)^q\lambda'(z)}\notag \\
=&-\big(\res_{z=z_i}+\res_{z=z_j}\big)\left(\frac{\lambda''(z)}{(z_i-z)^p(z_j-z)^q\lambda'(z)}\right).
\end{align}
Here the positive integers $p, q$ satisfy $p+q\ge 3$.
From these identities and the ones given in \eqref{l5}--\eqref{l7} it follows that
\begin{align}
P_{ij}^{(2)}=&h_i^2h_j^2X+\frac{7}{2880z_{ij}^2}\left(\sum_{k\neq i}\frac{2h_j^2h_k^2C_{k3}}{z_{ik}^3}-\sum_{k\neq j}\frac{h_i^2h_k^2(C_{k3}^2-C_{k4})}{z_{jk}^2}\right) \notag \\
=&h_i^2h_j^2X+\frac{7}{2880z_{ij}^2}\left(\sum_{k\neq i}\frac{2h_j^2h_k^2C_{k3}}{z_{ik}^3}-\sum_{k\neq j}\frac{2h_i^2h_k^2C_{k3}}{z_{jk}^3}\right) \notag \\
&-\frac{7h_i^2}{2880z_{ij}^2}\sum_{k\neq j}\frac{h_k^2}{z_{jk}^2}\left(C_{k3}^2-C_{k4}-\frac{2C_{k3}}{z_{jk}}\right),
\end{align}
where  $X\in \mathbb{Q}[\dfrac{1}{z_{ij}},C_{i3},C_{j3},C_{i4},C_{j4},C_{i5},C_{j5},C_{i6},C_{j6}]$.
By using the identity \eqref{iden-1} that is given in Lemma \ref{lem-2}, we can simplify $P_{ij}^{(2)}$ further to get the expression
\begin{align*}
P_{ij}^{(2)}
=&h_i^2 h_j^2 X+\frac{7}{2880z_{ij}^2}\left(\sum_{k\neq i}\frac{2h_j^2h_k^2C_{k3}}{z_{ik}^3}-\sum_{k\neq j}\frac{2h_i^2h_k^2C_{k3}}{z_{jk}^3}\right)\\
&-\frac{7h_i^2}{2880z_{ij}^2}\left(-\frac{1}{2}C_{j3}^4+\frac{5}{4}C_{j3}^2C_{j4}-\frac{1}{3}C_{j4}^2-\frac{7}{12}C_{j3}C_{j5}+\frac{1}{6}C_{j6}\right)\\
=&\frac{7}{2880z_{ij}^2}\left(\sum_{k\neq i}\frac{2h_j^2h_k^2C_{k3}}{z_{ik}^3}-\sum_{k\neq j}\frac{2h_i^2h_k^2C_{k3}}{z_{jk}^3}\right)+Y_{ij}-Y_{ji},\\
\end{align*}
where
\begin{align*}
Y_{ij}=&\frac{h_i^2 h_j^2}{5760z_{ij}^2}\left(-\frac{22C_{i3}}{z_{ij}^3}+\frac{19C_{i3}^2-104C_{i4}}{2z_{ij}^2}+\frac{15C_{i3}^3-34C_{i3}C_{i4}+21C_{i5}}{z_{ij}}\right.\\
&\left.+\frac{45}{4}C_{i3}^4
-22C_{i3}^2C_{i4}-\frac{1}{6}C_{j3}^2C_{i4}+5C_{i4}^2+\frac{49}{6}C_{i3}C_{i5}-\frac{23}{10}C_{i6}\right)
\end{align*}
Thus we proved the expected equation $P_{ij}^{(2)}=-P_{ji}^{(2)}$ for $i\ne j$.

\section{Proof of Theorem \ref{mth}}\label{sec-6}
In this section let us finish the proof of Theorem \ref{mth}. To this end we are left to prove that 
\begin{equation}\label{zh-5-1}
\frac{1}{2}P_{ii}^{(2)}+Q_i^{(2)}=0,\quad i=1,\dots,n.
\end{equation} 

From Lemma \ref{zh-16} it follows that the first term in the above mentioned summation has the expression 
\begin{equation}\label{zh-30-1}
\frac{1}{2}P_{ii}^{(2)}=\frac{h_i^4}{480}\sum_{k\neq i}\left(\frac{1}{z_{ik}^6}+\frac{C_{k3}}{2z_{ik}^5}\right).
\end{equation}
To prove the vanishing of \eqref{zh-5-1}, we need to calculate the 
summations w.r.t. the indices $k, l$ that appear in the expression of $Q_i^{(2)}$ 
defined in Append A. Such summations are difficult to calculate when their summands 
have denominators involving explicitly the difference of canonical coordinates $u_{ik}=u_i-u_k$. Due to this reason let us first pick up all such terms from $Q_i^{(2)}$ and denote the sum of them by $\al_1$, i.e.
\begin{align}\alpha_1:=&\sum_{k\ne i}\left(\frac{\gamma_{ik}H_k}{576u_{ik}h_ih_k}+\frac{\gamma_{ik}h_kH_i}{576u_{ik}h_i^3}
-\frac{\partial_i\gamma_{ik}h_k}{576u_{ik}h_i^3}-\frac{\partial_k\gamma_{ik}}{576u_{ik}h_ih_k}\right)\notag\\
&+\sum_{k,l\ne i}\left(\frac{u_{lk}\gamma_{ik}\partial_l\gamma_{kl}}{1152u_{il}h_ih_l}
+\frac{h_ku_{kl}\gamma_{kl}\partial_i\gamma_{il}}{1152u_{ik}h_i^3}\right).\label{zh-5-2}\end{align}
By using Lemma \ref{zh-16}, we can rewrite the last two terms in the above expression of $\alpha_1$ in the following form:
\begin{align}
&\sum_{k,l\ne i}\frac{u_{lk}\gamma_{ik}\partial_l\gamma_{kl}}{1152u_{il}h_ih_l}
=\sum_{k,l\ne i}\frac{u_{kl}\gamma_{il}\partial_k\gamma_{kl}}{1152u_{ik}h_ih_k}\notag\\
&\qquad =\sum_{k\neq i}\frac{h_k^2}{1152u_{ik}}\sum_{l\neq k,i}\frac{u_{kl}h_l^2}{z_{il}^2z_{kl}^2}
\left(\frac{C_{k3}^2}{4}-\frac{C_{k4}}{6}+\frac{3}{z_{kl}^2}+\frac{C_{k3}}{z_{kl}}\right)\label{zh-5-3}\\
&\sum_{k,l\ne i}\frac{h_ku_{kl}\gamma_{kl}\partial_i\gamma_{il}}{1152u_{ik}h_i^3}\notag\\
&\qquad =
\sum_{k\neq i}\frac{h_k^2}{1152u_{ik}}\sum_{l\neq k,i}\frac{u_{kl}h_l^2}{z_{il}^2z_{kl}^2}
\left(\frac{C_{i3}^2}{4}-\frac{C_{i4}}{6}+\frac{3}{z_{il}^2}+\frac{C_{i3}}{z_{il}}\right).\label{zh-5-4}
\end{align}
In the r.h.s. of the above formulae the summation w.r.t. $l$ can be represented as\begin{displaymath}\begin{split}
\sum_{l\neq k,i}\frac{u_{kl}h_l^2}{z_{il}^2 z_{kl}^2}
=&\sum_{l\neq k,i}\res_{z=z_l}\frac{\lambda(z_k)-\lambda(z)}{(z_i-z)^2 (z_k-z)^2\lambda'(z)}\\
=&-\big(\res_{z=z_i}+\res_{z=z_k}\big)\left(\frac{\lambda(z_k)-\lambda(z)}{(z_i-z)^2 (z_k-z)^2 \lambda'(z)}\right).
\end{split}\end{displaymath}
Then by substituting the r.h.s. of the residue formula for $R_6(2,2)$ given in Append B 
into \eqref{zh-5-3} and \eqref{zh-5-4} and by using Lemma \ref{zh-16}, we can represent the sum $\al_1$ in terms of $z_{kl}, C_{kl}, h_k$ and $u_{k}$. Note that all terms that contain $u_{ik}$ in the denominators are cancelled.

Let us continue to exam summations of other terms in $Q_i^{(2)}$. We denote by $\alpha_2$ the terms in $Q_i^{(2)}-\alpha_1$ given by the double summation w.r.t. to 
the indices $k, l$. By using Lemma \ref{zh-16} we have
\begin{align}
\alpha_2=&\sum_{k\neq i}\left(\frac{h_i^2h_k^2}{z_{ik}^2}\sum_{l\neq i}\left(\frac{h_l^2u_{il}}{40z_{il}^4}
\left(\frac{C_{i3}^2}{4}-\frac{C_{i4}}{6}+\frac{3}{z_{ik}^2}+\frac{C_{i3}}{z_{ik}}\right)\right.\right.\notag\\
&\qquad -\frac{1}{1440z_{il}^2}\left(\left.\frac{3}{z_{ik}^2}-\frac{C_{i3}}{z_{ik}}-\frac{C_{i3}^2}{4}+\frac{C_{i4}}{6}\right)\right)\notag\\
&\qquad +\frac{h_i^2h_k^2}{144z_{ik}^2}\left(\frac{3}{z_{ik}^2}+\frac{C_{i3}}{z_{ik}}+\frac{C_{i3}^2}{4}-\frac{C_{i4}}{6}\right)\sum_{l\neq k,i}\frac{h_l^2u_{lk}}{z_{il}^2z_{kl}^2}\notag\\
&\qquad -\frac{h_i^2}{34560z_{ik}^2}\Big(\frac{36h_i^2+24h_k^2}{z_{ik}^2}+\frac{C_{i3}(12h_i^2+12h_k^2)}{z_{ik}}\notag\\
&\qquad \qquad +\left.(3C_{i3}^2-2C_{i4})(2h_i^2+h_k^2)\Big)\sum_{l\neq k}\frac{1}{z_{kl}^2}\right).\label{zh-5-5}
\end{align}
By using the residue formulae for $R_3(4)$ and $R_6(2,2)$ given in Appendix B we have
\begin{displaymath}\begin{split}
\sum_{l\neq i}\frac{h_l^2u_{il}}{z_{il}^4}=&\sum_{l\neq i}\res_{z=z_l}\frac{\lambda(z_i)-\lambda(z)}{(z-z_i)^4\lambda'(z)}
=-\res_{z=z_i}\frac{\lambda(z_i)-\lambda(z)}{(z-z_i)^4\lambda'(z)}\\
=&\frac{1}{24}\big(C_{i3}^2-C_{i4}\big)\\
\sum_{l\neq k,i}\frac{h_l^2u_{lk}}{z_{il}^2z_{kl}^2}
=&\sum_{l\neq k,i}\res_{z=z_l}\frac{\lambda(z)-\lambda(z_k)}{(z_i-z)^2(z_k-z)^2\lambda'(z)}\\
=&-\big(\res_{z=z_k}+\res_{z=z_i}\big)\frac{\lambda(z)-\lambda(z_k)}{(z_i-z)^2(z_k-z)^2\lambda'(z)}\\
=&-\frac{h_i^2u_{ik}}{z_{ik}^2}\Big(\frac{3}{z_{ik}^2}+\frac{C_{i3}}{z_{ik}}+\frac{C_{i3}^2}{4}-\frac{C_{i4}}{6}\Big)
-\frac{1}{z_{ik}^2}
\end{split}\end{displaymath}
Substituting these expressions of the summation w.r.t. the index $l$ into the r.h.s. of 
\eqref{zh-5-5}, we obtained an expression of $\al_{2}$ without summations w.r.t.
the index $l$.

Let us denote by $\beta_0$ the summation of the first four terms in the expression of $Q_i^{(2)}$, i.e.
\[\beta_0=\frac{4H_i^3}{5h_i^2}-\frac{7\partial_ih_iH_i^2}{10h_i^3}+\frac{7(\partial_ih_i)^2H_i}{48h_i^4}
-\frac{(\partial_ih_i)^3}{120h_i^5},\]
and we denote $\beta_1=\frac{1}{2}P_{ii}^{(2)}+Q_i^{(2)}-\beta_0$. Then it follows from the result of our calculation for $\al_1, \al_2$ that 
\begin{align*}
\beta_1=&\sum_{k\neq i}\left(-\frac{h_i^4h_k^2u_{ik}}{16z_{ik}^8}-\frac{h_i^4C_{i3}h_k^2u_{ik}}{24z_{ik}^7}
-\frac{41h_i^2h_k^2}{960z_{ik}^6}+\frac{h_i^4}{120z_{ik}^6}-\frac{h_i^4(25C_{i3}^2-10C_{i4})h_k^2u_{ik}}{1440z_{ik}^6}\right.\\
&+\frac{h_i^4C_{i3}}{480z_{ik}^5}+\frac{h_i^4C_{k3}}{960z_{ik}^5}-\frac{73h_i^2C_{i3}h_k^2}{5760z_{ik}^5}
-\frac{h_i^4(15C_{i3}^2-10C_{i4})h_k^2u_{ik}}{4320z_{ik}^5}+\frac{h_i^4(3C_{i3}^2-4C_{i4})}{11520z_{ik}^4}\\
&+\frac{h_i^4C_{k4}}{5760z_{ik}^4}+\frac{h_i^2(53C_{i3}^2+38C_{i4})h_k^2}{69120z_{ik}^4}-\frac{h_i^2(3C_{i3}^2-2C_{i4})h_k^2u_{ik}}{20736z_{ik}^4}-\frac{h_i^4C_{i3}^3}{11520z_{ik}^3}\\
&+\frac{h_i^2(17C_{i3}^3-32C_{i3}C_{i4}+5C_{i5})h_k^2}{20340z_{ik}^3}+\frac{h_i^4C_{i3}C_{k4}}{17280z_{ik}^3}+\frac{h_i^4(9C_{i3}^4-12C_{i3}^2C_{i4}+4C_{i4}^2)}{414720z_{ik}^2}\\
&\left.+\frac{h_i^4(3C_{i3}^2-2C_{i4})C_{k4}}{103680z_{ik}^2}+\frac{h_i^2(153C_{i3}^4-390C_{i3}^2C_{i4}+92C_{i4}^2+120C_{i3}C_{i5}-18C_{i6})h_k^2}{829440z_{ik}^2}\right)\\
&-\frac{h_i^2}{4608}\left(\frac{C_{i3}^2}{5}+\frac{C_{i4}}{3}\right)\sum_{k\neq i}\frac{h_k^2}{z_{ik}^2}\left(C_{k3}^2-C_{k4}-\frac{2C_{k3}}{z_{ik}}\right)\\
&+\frac{h_i^2C_{i3}}{2304}\sum_{k\neq i}\frac{h_k^2}{z_{ik}^3}\left(C_{k3}^2-C_{k4}-\frac{3C_{k3}}{z_{ik}}\right)
+\frac{17h_i^2}{11520}\sum_{k\neq i}\frac{h_k^2}{z_{ik}^4}\left(C_{k3}^2-C_{k4}-\frac{4C_{k3}}{z_{ik}}\right).
\end{align*}
By using the identities \eqref{l5}\eqref{l6}\eqref{l7}, the identity 
\begin{align}
&\sum_{k\neq i}\frac{h_k^2u_{ik}}{z_{ik}^p}=\sum_{k\neq i}\res_{z=z_k}\frac{\lambda(z_i)-\lambda(z)}{(z_i-z)^p\lambda'(z)}=-\res_{z=z_i}\frac{\lambda(z_i)-\lambda(z)}{(z_i-z)^p\lambda'(z)},
\end{align}
and the residue formulae for $R_1(p), R_2(p), R_3(p), R_4(p, q)$ we obtain
\begin{align}
&\frac{1}{2}P_{ii}^{(2)}+Q_i^{(2)}\label{zh-5-6}\\
&=h_i^4 Z-\frac{h_i^2}{4608}\left(\frac{C_{i3}^2}{5}+\frac{C_{i4}}{3}\right)\sum_{k\neq i}\frac{h_k^2}{z_{ik}^2}\left(C_{k3}^2-C_{k4}-\frac{2C_{k3}}{z_{ik}}\right)\notag \\
&\quad+\frac{h_i^2C_{i3}}{2304}\sum_{k\neq i}\frac{h_k^2}{z_{ik}^3}\left(C_{k3}^2-C_{k4}-\frac{3C_{k3}}{z_{ik}}\right)
+\frac{17h_i^2}{11520}\sum_{k\neq i}\frac{h_k^2}{z_{ik}^4}\left(C_{k3}^2-C_{k4}-\frac{4C_{k3}}{z_{ik}}\right).\notag
\end{align}
Here $Z\in \mathbb{Q}[C_{i3},C_{i4},C_{i5},C_{i6},C_{i7},C_{i8}]$.
We can simplify the above expression further to get
\begin{align*}
&\frac{1}{2}P_{ii}^{(2)}+Q_i^{(2)}\\
&=\frac{h_i^2C_{i3}}{2304}\sum_{k\neq i}\frac{h_k^2}{z_{ik}^3}\left(C_{k3}^2-C_{k4}-\frac{3C_{k3}}{z_{ik}}\right)+\frac{17h_i^2}{11520}\sum_{k\neq i}\frac{h_k^2}{z_{ik}^4}\left(C_{k3}^2-C_{k4}-\frac{4C_{k3}}{z_{ik}}\right)\\
&\quad+\frac{h_i^4}{8294400}\left(3420C_{i3}^6-12525C_{i3}^4C_{i4}+6390C_{i3}^3C_{i5}+10260C_{i3}^2C_{i4}^2\right.\\
&\quad-2496C_{i3}^2C_{i6}-6000C_{i3}C_{i4}C_{i5}+694C_{i3}C_{i7}-1020C_{i4}^3\\
&\quad\left.+884C_{i4}C_{i6}+510 C_{i5}^2-102C_{i8}\right).
\end{align*}
Then the equality \eqref{zh-5-1} follows from the identities \eqref{iden-3},
\eqref{iden-4} given in Lemma \ref{lem-2}.

Finally, By combining the equality \eqref{zh-5-1} with the results of the previous sections on the vanishing of 
$G_i^{(2)}, G_{ij}^{(2)}$ and $P_{ij}+P_{ji}\ ( i\ne j)$ 
we arrive at the result of Theorem \ref{mth}, i.e. $G^{(2)}=0$. The Theorem is proved.

\section{Conclusion} \label{sec-7}
We proved in this paper the vanishing of the genus two G-functions for the Frobenius manifolds obtained from the simple singularities of type A by using the formulae 
of the rotation coefficients and the Lam\'e coefficients given in \cite{DLZ}. 
Similar formulae are also given in \cite{DLZ} for the Frobenius manifolds associated to
the simple singularities type D, and for the ones associated to the extended affine Weyl groups 
of type A and type D. We hope that the prove of the Conjecture \ref{main-conj1} can also 
be obtained for these classes of Frobenius manifolds by using a similar argument as given 
in the present paper. We will return to the proof of the conjecture in subsequent publications. 

\vskip0.6 truecm
\noindent{\bf Acknowledgments.}
\vskip0.3 truecm

The authors thank Boris Dubrovin for his encouragement and helpful discussions. This work is partially supported by the NSFC No.\,11071135, No.\,11171176
and No. 11222108, and by the Marie Curie IRSES project RIMMP.

\appendix
\setcounter{section}{0}

\section{The genus two $G$-function}\label{app1}

We recalled the following expression of the genus two $G$-function $G^{(2)}(u, u_x, u_{xx})$ that is given in the appendix of \cite{DLZ}:
\begin{align}\label{g2-zh}
G^{(2)}(u, u_x, u_{xx})=&\sum_{i=1}^n G^{(2)}_i(u,u_x)u^i_{xx}+\sum_{i\ne j} G^{(2)}_{ij}(u) \frac{(u^j_x)^3}{u^i_x}\nn\\
&+ {\frac12\sum_{i,j}P^{(2)}_{ij}(u) u^i_x u^j_x}+\sum_{i=1}^n Q_i^{(2)} (u) \left(u_x^i\right)^2
\end{align}
with coefficients written in terms of the Lam\'e coefficients $h_i=h_i(u)$ and rotation coefficients $\gamma_{ij}=\gamma_{ij}(u)$ of the semisimple Frobenius manifold.  Here
\begin{align}
G^{(2)}_i=&\frac{\p_x h_i\,H_i}{60\, u_{i,x} h_i^3}-\frac{3\,\p_i h_i H_i}{40\,h_i^3}
+\frac{19\,(\p_i h_i)^2}{2880\,h_i^4}-\frac{7\,\p_i h_i \p_x h_i}{5760\,u_{i,x} h_i^4}\nn
\\
&+\sum _{k} \left[\frac{\gamma_{i k} H_i}{120\, h_i h_k}+
\frac{\gamma_{i k} H_k}{120\, h_i h_k}\left(7+\frac{u_{k,x}}{u_{i,x}}\right)
-\frac{\gamma_{i k}}{5760\,h_i^2 h_k}\left(4\,\p_i h_i+\frac{\p_x h_i}{u_{i,x}}\right)
\right.\nn\\
&\left.
-\frac{\gamma_{i k}\p_k h_k}{h_i h_k^2}\left(\frac{u_{k,x}}{1152\,u_{i,x}}+
\frac{7}{2880}\right)+\frac{\gamma_{i k} \p_k h_k}{384\,h_i^3}
-\frac{\p_k \gamma_{i k} h_k}{384\,h_i^3}
+\frac{\p_i \gamma_{i k}h_k u_{k,x}}{1920\,u_{i,x}h_i^3}\right.\nn\\
&\left.
+\frac{\p_i\gamma_{i k}}{2880\,h_i h_k}
+\frac{\p_x\gamma_{i k}}{5760\,u_{i,x}h_i h_k}
+\frac{\p_k\gamma_{i k}}{h_i h_k}\left(\frac{u_x^k}{2880\, u_x^i}+\frac{7}{2880}\right)+\frac{\gamma_{i k} h_i \p_k h_k}{2880\,h_k^4}
\right.\nn\\
&\left.-\frac{\gamma_{ik}^2}{h_i^2}\left(\frac{7 u_x^k}{1152\, u_x^i}+\frac{19}{720}\right)+\frac{\gamma_{ik}^2}{1440\, h_k^2}\right]-\sum _{k, l} \left(\frac{h_i \gamma_{il} \gamma_{kl}}{2880\, h_k h_l^2}+\frac{u_{k,x}h_k \gamma_{il} \gamma_{kl}}{1920\, u_x^i h_i h_l^2}\right),\nn
\end{align}

\begin{align*}
G^{(2)}_{ij}=&
-\frac{\gamma_{i j}^2 H_j}{120\, h_j^2}
+\frac{\gamma _{i j}^3}{480\, h_i h_j}
-\frac{\gamma _{i j}}{5760}\left(\frac{\p_i\gamma_{i j}}{h_i^2}+\frac{\p_j\gamma_{i j}}{h_j^2}\right)
+\frac{\gamma _{i j}^2}{5760}\left(\frac{\p_i h_{i}}{h_i^3}+\frac{3\,\p_j h_{j}}{h_j^3}
\right)\\
&+\sum_k \left(
\frac{\gamma_{i j}\gamma_{i k}\gamma_{j k}}{5760\, h_k^2}+\frac{\gamma_{i j}^2}{5760\,h_k}\left(\frac{\gamma_{j k}}{h_j}-\frac{\gamma_{i k}}{h_i}\right)
\right),
\end{align*}

\begin{align*}
P^{(2)}_{ij}&=-\frac{2\, \gamma_{i j} H_i H_j}{5\,h_i h_j}
+\frac{\gamma_{i j} \p_j h_j H_i}{20\, h_i h_j^2}
+\frac{\gamma_{i j} h_i \p_j h_j H_j}{20\,h_j^4}
-\frac{19\,\gamma_{i j}^2 H_j}{30\,h_j^2}
-\frac{\p_i\gamma_{i j} H_j}{60\,h_i h_j}\\
&
+\frac{41\,\gamma_{i j}^3}{240\,h_i h_j}
-\frac{41 \gamma _{i j} \partial_i\gamma_{ij}}{1440\, h_i^2 }
+\frac{\p_i\gamma_{i j}\p_j h_j}{1440\,h_i h_j^2}
+\frac{79\,\gamma_{i j}^2 \p_j h_j}{1440\,h_j^3}
-\frac{\gamma_{i j} \p_i h_i \p_j h_j}{720\, h_i^2 h_j^2}
-\frac{\gamma_{i j} h_i (\p_j h_j)^2}{288\,h_j^5}\\
&+\sum_k\left(
\frac{\gamma_{i j}\gamma_{i k} H_j}{60\,h_j h_k}
-\frac{\gamma_{i k}\gamma_{j k} h_i h_j H_k}{30\, h_k^4}
-\frac{\gamma_{i j}\gamma_{j k} h_i H_j}{60\, h_j^2 h_k}
+\frac{\gamma_{i k}\gamma_{j k} h_i H_j}{60\, h_j h_k^2}
-\frac{7\,\gamma_{i j}\gamma_{j k} h_i H_k}{60\, h_j^2 h_k}\right.\\
&\qquad\left.
-\frac{\gamma_{i j}\gamma_{i k} \p_j h_j}{720\, h_j^2 h_k}
+\frac{\gamma_{i j}\gamma_{j k} h_i  \p_j h_j}{240\, h_j^3 h_k}
-\frac{\gamma_{i k}\gamma_{j k} h_i  \p_j h_j}{1440\, h_j^2 h_k^2}
+\frac{\gamma_{i j}\gamma_{j k} h_i  \p_k h_k}{720\, h_k^4}
+\frac{\gamma_{i k}\gamma_{j k} h_i h_j  \p_k h_k}{288\, h_k^5}
\right.
\end{align*}
\begin{align*}
&\qquad\left.+\frac{ \gamma _{j k}\,\partial_i\gamma _{i j}}{1440\, h_i h_k}
-\frac{h_j h_k \gamma _{i j} \partial_i\gamma _{i k}}{360\, h_i^4 }
-\frac{h_j (3\,\gamma_{ik}\partial_i\gamma_{ij}+2\,\gamma_{i j} \p_i\gamma_{i k})}{1440\, h_i^2 h_k }
-\frac{7\,h_j \gamma _{i j} \partial_k (h_k^{-1}\gamma _{i k})}{1440\, h_i^2 }
\right.\\
&\qquad\left.
-\frac{h_i h_j \gamma _{i k}\, \partial_k\gamma _{j k}}{480 \,h_k^4 }+\frac{\gamma _{i j}^2 \gamma _{j k}}{120\, h_j h_k}
+\frac{7\, h_i \gamma _{i j} \gamma _{j k}^2}{160 \,h_j^3}
+\frac{11 \gamma _{i j} \gamma _{i k} \gamma _{j k}}{2880\, h_k^2}+\frac{h_j \gamma _{i k}^2 \gamma _{j k}}{96 \,h_k^3}\right)\\
&+\sum_{k, l}\left(
\frac{h_i h_j \gamma _{i l} \gamma _{j l} }{720\, h_k h_l^2}
\left(\frac{\gamma _{k l}}{h_l}-\frac{\gamma_{j k}}{2 h_j}\right)
-\frac{h_i  \gamma_{i j}\gamma _{j l} \gamma _{k l} }{720\,h_k h_l^2}
\right),
\end{align*}

\begin{align*}
Q^{(2)}_{i}
=&\frac{4 H_i^3}{5\, h_i^2}
-\frac{7\, \p_i h_i H_i^2}{10\,h_i^3}
+\frac{7\, (\p_i h_i)^2 H_i}{48\, h_i^4}-\frac{(\p_i h_i)^3}{120\,h_i^5}+\sum_k \left(\frac{7 \gamma_{i k} H_i H_k }{10\, h_i h_k}
-\frac{\gamma_{i k}\p_i h_i H_i}{120\, h_i^2 h_k}\right.\\
&\left.
+\frac{7 \,\p_k\left(h_k^{-1} \gamma_{i k}\right)H_i}{240\, h_i }
-\frac{7  \gamma_{i k} \p_i h_i H_k}{80\, h_i^2 h_k}
+\frac{\gamma_{i k}H_k}{576\, u_{i k} h_i h_k}
+\frac{(2 H_i+7 H_k)\p_i\gamma_{i k}}{240\,h_i h_k}\right.\\
&\left.
+\frac{\gamma_{i k} h_k H_i}{576\, u_{i k} h_i^3}
-\frac{31  \gamma_{i k}^2 H_i}{144\, h_i^2}
+\frac{\gamma_{i k} (\p_i h_i)^2}{720\, h_i^3 h_k}
+\frac{253\,\gamma_{i k}^2\p_i h_i }{5760\,h_i^3}
-\frac{\p_i\gamma_{i k} \p_i h_i }{960\,h_i^2 h_k}
-\frac{\gamma_{i k}^2 \p_k h_k}{2880 \,h_k^3}\right.\\
&\left.-\frac{7\, \p_k\left(h_k^{-1}\gamma_{i k}\right)\p_i h_i}{1920\, h_i^2 }
-\frac{7\,\p_i\gamma_{i k}\p_k h_k}{5760\,h_i h_k^2}
-\frac{41\,\p_i\gamma_{i k} \p_i h_i h_k}{5760\,h_i^4}
+\frac{ \p_i (h_i\gamma_{i k}) \p_k h_k}{2880\, h_k^4 }\right.\\
&\left.
-\frac{113\, \gamma _{i k}\partial_i\gamma_{i k}}{5760\, h_i^2}
+\frac{\left(3\,\partial_i\gamma_{i k}+\partial_k\gamma_{i k}\right) \gamma _{i k}}{1440\, h_k^2 }
-\frac{ \partial_i \gamma _{i k}h_k}{576\, u_{i k} h_i^3 }
-\frac{\partial_k \gamma _{i k}}{576\, u_{i k} h_i h_k }
-\frac{\gamma _{i k}^3}{240\, h_i h_k}\right)\\
&+\sum_{k, l}  \left(-\frac{\gamma_{k l}\partial_i (h_i \gamma _{i l})}{2880\, h_k h_l^2 }
+\frac{\gamma _{i l}^2 \gamma _{k l}}{2880 \,h_k h_l}
-\frac{\gamma _{i k} \gamma _{i l}^2}{240\,h_i h_k }
-\frac{\gamma_{k l}\partial_i\gamma _{i k} }{2880 \,h_i h_l }
+\frac{u_{l k}\gamma _{i k}\partial_l \gamma _{k l} }{1152\,u_{i l} h_i h_l }
\right.\\
&\left.
+\frac{ u_{k l}\gamma _{i k} \gamma_{k l} \partial_i\gamma _{i l} }{144\,  h_i^2}
+\frac{ h_l \gamma _{i k}\partial_i \gamma _{i l}}{1440 \, h_i^2 h_k}
+\frac{h_k u_{k l} \gamma_{k l} \partial_i \gamma _{i l} }{1152\, u_{i k} h_i^3}
+\frac{h_l u_{i k}\gamma _{i k}^2 \partial_i\gamma _{i l}}{40\,h_i^3}
\right).
\end{align*}

In these expressions, the summations are taken over indices such that the denominators do not vanish.

\section{Some residue formulae}
Here is the list of most of the residue formulae that we used in this paper.

\noindent Residues of form $R_1(p):=\res_{z=z_i}\frac{\lambda''(z)}{(z_i-z)^p\lambda'(z)}$\,.
\begin{align*}
&R_1(2)=-\frac{C_{i3}^2}{4}+\frac{C_{i4}}{3};\\
&R_1(3)=-\frac{1}{8}\left(C_{i3}^3-2C_{i3}C_{i4}+C_{i5}\right);\\
&R_1(4)=-\frac{1}{720}\left(45C_{i3}^4-120C_{i3}^2C_{i4}+40C_{i4}^2+60C_{i3}C_{i5}-24C_{i6}\right);\\
&R_1(5)=-\frac{1}{288}\left(9C_{i3}^5-30C_{i3}^3C_{i4}+20C_{i3}C_{i4}^2+15C_{i3}^2C_{i5}-10C_{i4}C_{i5}-6C_{i3}C_{i6}+2C_{i7}\right);
\end{align*}
\begin{align*}
&R_1(6)=-\frac{1}{60480}\left(945C_{i3}^6-3780C_{i3}^4C_{i4}+3780C_{i3}^2C_{i4}^2-560C_{i4}^3+1890C_{i3}^3C_{i5};\right.\\
&\qquad\quad\ \ \left.-2520C_{i3}C_{i4}C_{i5}+315C_{i5}^2-756C_{i3}^2C_{i6}+504C_{i4}C_{i6}+252C_{i3}C_{i7}-72C_{i8}\right).
\end{align*}
Residues of form $R_2(p)=\res_{z=z_i}\frac{1}{(z_i-z)^p\lambda'(z)}$\,.
\begin{align*}
&R_2(2)=\frac{h_i^2}{12}\left(3C_{i3}^2-2C_{i4}\right);\\
&R_2(3)=\frac{h_i^2}{24}\left(3C_{i3}^3-4C_{i3}C_{i4}+C_{i5}\right);\\
&R_2(4)=\frac{h_i^2}{720}\left(45C_{i3}^4-90C_{i3}^2C_{i4}+20C_{i4}^2+30C_{i3}C_{i5}-6C_{i6}\right);\\
&R_2(5)=\frac{h_i^2}{1440}\left(45C_{i3}^5-120C_{i3}^3C_{i4}+60C_{i3}C_{i4}^2+45C_{i3}^2C_{i5}-20C_{i4}C_{i5}\right.\\
&\qquad\quad \ \left. -12C_{i3}C_{i6}+2C_{i7}\right);\\
&R_2(6)=\frac{h_i^2}{60480}\left(945C_{i3}^6-3150C_{i3}^4C_{i4}+2520C_{i3}^2C_{i4}^2-280C_{i4}^3+1260C_{i3}^3C_{i5}\right.\\
&\qquad\quad\ \left.-1260C_{i3}C_{i4}C_{i5}+105C_{i5}^2-378C_{i3}^2C_{i6}+168C_{i4}C_{i6}+84C_{i3}C_{i7}-12C_{i8}\right).
\end{align*}
Residues of form $R_3(p):=\res_{z=z_i}\frac{\lambda(z_i)-\lambda(z)}{(z_i-z)^p\lambda'(z)}$\,.
\begin{align*}
&R_3(4)=-\frac{1}{24}\left(C_{i3}^2-C_{i4}\right);\\
&R_3(5)=-\frac{1}{720}\left(15C_{i3}^3-25C_{i3}C_{i4}+9C_{i5}\right);\\
&R_3(6)=-\frac{1}{1440}\left(15C_{i3}^4-35C_{i3}^2C_{i4}+10C_{i4}^2+14C_{i3}C_{i5}-4C_{i6}\right);\\
&R_3(7)=-\frac{1}{60480}\left(315C_{i3}^5-945C_{i3}^3C_{i4}+560C_{i3}C_{i4}^2+399C_{i3}^2C_{i5}-231C_{i4}C_{i5}\right.\\
&\qquad\quad\  \left.-126C_{i3}C_{i6}+30C_{i7}\right);\\
&R_3(8)=-\frac{1}{120960}\left(315C_{i3}^6-1155C_{i3}^4C_{i4}+1050C_{i3}^2C_{i4}^2-140C_{i4}^3+504C_{i3}^3C_{i5}\right.\\
&\qquad\quad \ \left.-602C_{i3}C_{i4}C_{i5}+63C_{i5}^2-168C_{i3}^2C_{i6}+98C_{i4}C_{i6}+44C_{i3}C_{i7}-9C_{i8}\right).
\end{align*}
Residues of form $R_4(p,q):=\res_{z=z_i}\frac{\lambda^{(q)}(z)}{(z_i-z)^p\lambda'(z)}$\,.
\begin{align*}
&R_4(5,3)=\frac{1}{1440}\left(45C_{i3}^6-210C_{i3}^4C_{i4}+240C_{i3}^2C_{i4}^2-40C_{i4}^3+135C_{i3}^3C_{i5}\right.\\
&\qquad\qquad\  \left.-200C_{i3}C_{i4}C_{i5}+30C_{i5}^2-72C_{i3}^2C_{i6}+52C_{i4}C_{i6}+32C_{i3}C_{i7}-12C_{i8}\right);\\
&R_4(2,4)=\frac{1}{12}\left(3C_{i3}^2C_{i4}-2C_{i4}^2-6C_{i3}C_{i5}+6C_{i6}\right);\\
&R_4(3,4)=\frac{1}{24}\left(3C_{i3}^3C_{i4}-4C_{i3}C_{i4}^2-6C_{i3}^2C_{i5}+5C_{i4}C_{i5}+6C_{i3}C_{i6}-4C_{i7}\right);\\
&R_4(4,4)=\frac{1}{720}\left(45C_{i3}^4C_{i4}-90C_{i3}^2C_{i4}^2+20C_{i4}^3-90C_{i3}^3C_{i5}\right.\\
&\qquad\qquad\ \left.+150C_{i3}C_{i4}C_{i5}-30C_{i5}^2+90C_{i3}^2C_{i6}-66C_{i4}C_{i6}-60C_{i3}C_{i7}+30C_{i8}\right).
\end{align*}
Residues of form $R_5(p,q):=\left(\res_{z=z_i}+\res_{z=z_j}\right)\left(\frac{\lambda''(z)}{(z_i-z)^p(z_j-z)^q\lambda'(z)}\right)$\,.
\begin{align*}
&R_5(2,2)=\dfrac{6}{z_{ij}^4}-\dfrac{C_{i3}-C_{j3}}{z_{ij}^3}-\dfrac{3C_{i3}^2+3C_{j3}^2-4C_{i4}-4C_{j4}}{12z_{ij}^2};\\
&R_5(2,4)=\dfrac{15}{z_{ij}^6}-\dfrac{2C_{i3}-2C_{j3}}{z_{ij}^5}-\dfrac{3C_{i3}^2+9C_{j3}^2-4C_{i4}-12C_{j4}}{12z_{ij}^4}+\dfrac{C_{j3}^3-2C_{j3}C_{j4}+C_{j5}}{4z_{ij}^3}\\
&\qquad\qquad \ -\dfrac{45C_{j3}^4-120C_{j3}^2C_{j4}+40(C_{j4})^2+60C_{j3}C_{j5}-24C_{j6}}{720z_{ij}^2};\\
&R_5(4,2)=\dfrac{15}{z_{ij}^6}-\dfrac{2C_{i3}-2C_{j3}}{z_{ij}^5}-\dfrac{3C_{j3}^2+9C_{i3}^2-4C_{j4}-12C_{i4}}{12z_{ij}^4}-\dfrac{C_{i3}^3-2C_{i3}C_{i4}+C_{i5}}{4z_{ij}^3}\\
&\qquad\qquad \  -\dfrac{45C_{i3}^4-120C_{i3}^2C_{i4}+40C_{i4}^2+60C_{i3}C_{i5}-24C_{i6}}{720z_{ij}^2}.
\end{align*}
Residues of form $R_6(p,q):=\left(\res_{z=z_i}+\res_{z=z_k}\right)\left(\frac{\lambda(z_k)-\lambda(z)}{(z_i-z)^p(z_k-z)^q\lambda'(z)}\right)$\,.\begin{align*}
&R_6(2,2)=-\frac{1}{z_{ik}^2}-\frac{h_i^2u_{ik}}{z_{ik}^2}\left(\frac{3}{z_{ik}^2}+\frac{C_{i3}}{z_{ik}}+\frac{C_{i3}^2}{4}-\frac{C_{i4}}{6}\right);\\
&R_6(2,3)=\frac{3}{2z_{ik}^3}-\frac{C_{k3}}{12z_{ik}^2}+\frac{h_i^2u_{ik}}{z_{ik}^3}\left(\frac{6}{z_{ik}^2}+\frac{3C_{i3}}{2z_{ik}}+\frac{C_{i3}^2}{4}-\frac{C_{i4}}{6}\right);\\
&R_6(2,4)=-\frac{2}{z_{ik}^4}+\frac{C_{k3}}{6 z_{ik}^3}-\frac{C_{k3}^2-C_{k4}}{24z_{ik}^2}-\frac{h_i^2u_{ik}}{z_{ik}^4}\left(\frac{10}{z_{ik}^2}+\frac{2C_{i3}}{z_{ik}}+\frac{C_{i3}^2}{4}-\frac{C_{i4}}{6}\right);\\
&R_6(3,2)=-\frac{3}{2 z_{ik}^3}-\frac{C_{i3}}{12z_{ik}^2}-\frac{h_i^2u_{ik}}{z_{ik}^2}\left(\frac{4}{z_{ik}^3}+\frac{3C_{i3}}{2z_{ik}^2}+\frac{2}{z_{ik}}\left(\frac{C_{i3}^2}{4}-\frac{C_{i4}}{6}\right)\right.\\
&\qquad\qquad\ \left.
+\frac{3C_{i3}^3-4C_{i3}C_{i4}+C_{i5}}{24}\right);\\
&R_6(4,2)=-\frac{2}{z_{ik}^4}-\frac{C_{i3}}{6z_{ik}^3}-\frac{C_{i3}^2-C_{i4}}{24z_{ik}^2}-\frac{h_i^2u_{ik}}{z_{ik}^2}\left(\frac{5}{z_{ik}^4}+\frac{2C_{i3}}{z_{ik}^3}+\frac{3}{z_{ik}^2}\left(\frac{C_{i3}^2}{4}-\frac{C_{i4}}{6}\right)\right.\\
&\quad\left.+\frac{3C_{i3}^3-4C_{i3}C_{i4}+C_{i5}}{12z_ik}+\frac{45C_{i3}^4-90C_{i3}^2C_{i4}+20C_{i4}^2+30C_{i3}C_{i5}-6C_{i6}}{720}\right).
\end{align*}

\end{document}